\documentclass[%
 reprint,
superscriptaddress, showpacs,
 amsmath,amssymb,
 aps,
prl]{revtex4-1}

\usepackage{graphicx}
\usepackage{dcolumn}
\usepackage{bm}
\usepackage[ansinew]{inputenc}

\begin{document}

\title{Non-perturbative Interband Response of InSb\\Driven Off-resonantly by Few-cycle Electromagnetic Transients}

\author{F. Junginger}
\affiliation{Department of Physics and Center for Applied Photonics, University of Konstanz, Universitätsstraße 10, 78464 Konstanz, Germany}
\author{B. Mayer}
\affiliation{Department of Physics and Center for Applied Photonics, University of Konstanz, Universitätsstraße 10, 78464 Konstanz, Germany}
\author{C. Schmidt}
\affiliation{Department of Physics and Center for Applied Photonics, University of Konstanz, Universitätsstraße 10, 78464 Konstanz, Germany}
\author{O. Schubert}
\affiliation{Department of Physics and Center for Applied Photonics, University of Konstanz, Universitätsstraße 10, 78464 Konstanz, Germany}
\affiliation{Department of Physics, University of Regensburg, Universitätsstraße 31, 93053 Regensburg, Germany}
\author{S.~Mährlein}
\affiliation{Department of Physics and Center for Applied Photonics, University of Konstanz, Universitätsstraße 10, 78464 Konstanz, Germany}
\author{A. Leitenstorfer}
\affiliation{Department of Physics and Center for Applied Photonics, University of Konstanz, Universitätsstraße 10, 78464 Konstanz, Germany}
\author{R. Huber}
\affiliation{Department of Physics and Center for Applied Photonics, University of Konstanz, Universitätsstraße 10, 78464 Konstanz, Germany}
\affiliation{Department of Physics, University of Regensburg, Universitätsstraße 31, 93053 Regensburg, Germany}
\author{A. Pashkin}
\affiliation{Department of Physics and Center for Applied Photonics, University of Konstanz, Universitätsstraße 10, 78464 Konstanz, Germany}

\date{\today}

\begin{abstract}
Intense multi-THz pulses are used to study the coherent nonlinear response of bulk InSb by means of field-resolved four-wave mixing
spectroscopy. At amplitudes above 5 MV/cm the signals show a clear temporal substructure which is unexpected in perturbative nonlinear optics.
Simulations based on a two-level quantum system demonstrate that in spite of the strongly off-resonant character of the excitation the
high-field pulses drive the interband resonances into a non-perturbative regime of Rabi flopping.
\end{abstract}


\maketitle

Semiconductors form a uniquely well-defined laboratory to explore novel limits of nonlinear optics. A key parameter for interaction of a
coherent light field with an electronic transition is given by the Rabi frequency $\Omega_R = \mu E / \hbar$. $E$ is the electric field strength
and $\mu$ the transition dipole moment. When the dephasing rate is negligible compared to $\Omega_R$, coherent Rabi flopping governs the
dynamics of electronic systems \cite{Cohen:77}. If the detuning of the driving electromagnetic field is smaller than $\Omega_R$ the response of
a system cannot be described by the perturbative approach which usually depicts off-resonant nonlinear optics. This non-perturbative excitation
regime provides access to many fascinating quantum effects in semiconductors such as Rabi splitting, self-induced transparency and generation of
high harmonics \cite{Wegener:90,Cundiff:94,Giessen:98,Schuelzgen:99,Muecke:01,Carter:05,Choi:10}. In particular, sufficiently intense and
ultrashort laser pulses have been exploited to implement ultimate scenarios in which the duration of the light pulse, the Rabi cycle and the
oscillation period of the carrier wave all become comparable in size \cite{Tritschler:03}. Under such conditions, a detailed insight into the
nonlinear optical interaction calls for complete phase and amplitude resolution of all interacting light fields. However, in most of the cases,
the lack of phase-stable laser pulses and fast detectors does not allow for capturing sub-cycle polarization dynamics of a system.
\begin{figure}[b]
\includegraphics[angle=270, width=8.6cm]{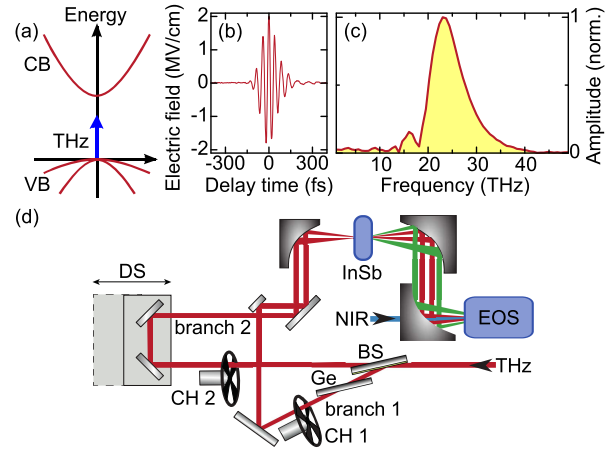}
\caption{\label{fig1} (a)~Schematic band structure of InSb with conduction (CB) and valence bands (VB). The blue arrow illustrates a
non-resonant excitation by a THz pulse. (b)~THz transient generated by a 370-$\mu$m-thick GaSe crystal. (c)~Amplitude spectrum of the THz
transient in b). (d)~Scheme of the FWM setup: The THz beam is split into two branches by a Ge beam splitter (BS). Both branches are modulated by
mechanical choppers (CH) and delayed with respect to each other by a translation stage (DS). The THz transients are non-collinearly focused on
the sample and the transmitted signals are detected by electro-optical sampling (EOS).}
\end{figure}

The development of ultraintense THz laser systems generating phase-stable transients with field amplitudes above 1 MV/cm
\cite{Reimann:03,Hebling:08,Hirori:11} paves the way towards a coherent spectroscopy of extreme nonlinearities in condensed matter systems with
absolute sampling of amplitude and phase. Recent experiments performed with high-field multi-THz pulses have demonstrated a high potential of
this approach \cite{Luo:04,Kuehn:09,Leinss:08,Kuehn:11}. However, the regime of an \emph{off-resonant} excitation of Rabi flopping has remained
almost unexplored due to the lack of sufficiently intense and phase-locked pulses. The latest breakthrough in generation and field-resolved
detection of multi-THz pulses with peak fields up to 100~MV/cm \cite{Sell:08,Junginger:10} opens up the possibility to explore this highly
non-perturbative regime.

In this Letter, we report the nonlinear response of the bulk semiconductor indium antimonide (InSb) excited far below interband resonance using
our novel high-field multi-THz laser source \cite{Junginger:10}:  Time-resolved four-wave mixing (FWM) signals are recorded with amplitude and
phase at different THz peak fields of the excitation pulses. For the highest intensities, the Rabi frequency becomes comparable to the detuning
and the interband resonance is driven into a non-perturbative regime of Rabi flopping. Our simulations of the FWM response based on the
Maxwell-Bloch equations provide a qualitative understanding of the phenomena observed experimentally.

The sample under study is a (100)-oriented undoped single crystal of InSb (Fig.~1(a)) mechanically polished to a thickness of 30~$\mu$m and kept
at room temperature. THz transients with a center frequency of $f_0 = 23$~THz, a bandwidth of 8~THz (full width at half maximum, FWHM) and
variable peak fields between 2~MV/cm and 5.3~MV/cm are generated by difference frequency mixing of near-infrared pulse trains with a repetition
rate of 1~kHz in gallium selenide (GaSe) emitters (Figs.~1(b) and 1(c)) \cite{Junginger:10}. A specific feature of our experiment is a large
detuning of 18~THz between the THz transients and the nearest interband resonance of InSb (E$_{g}/h = 41.1$~THz). Thus, the entire THz spectrum
is located well below the band gap excluding the possibility of direct linear excitation of electron-hole pairs.
\begin{figure}[b]
\includegraphics[width=8.6cm]{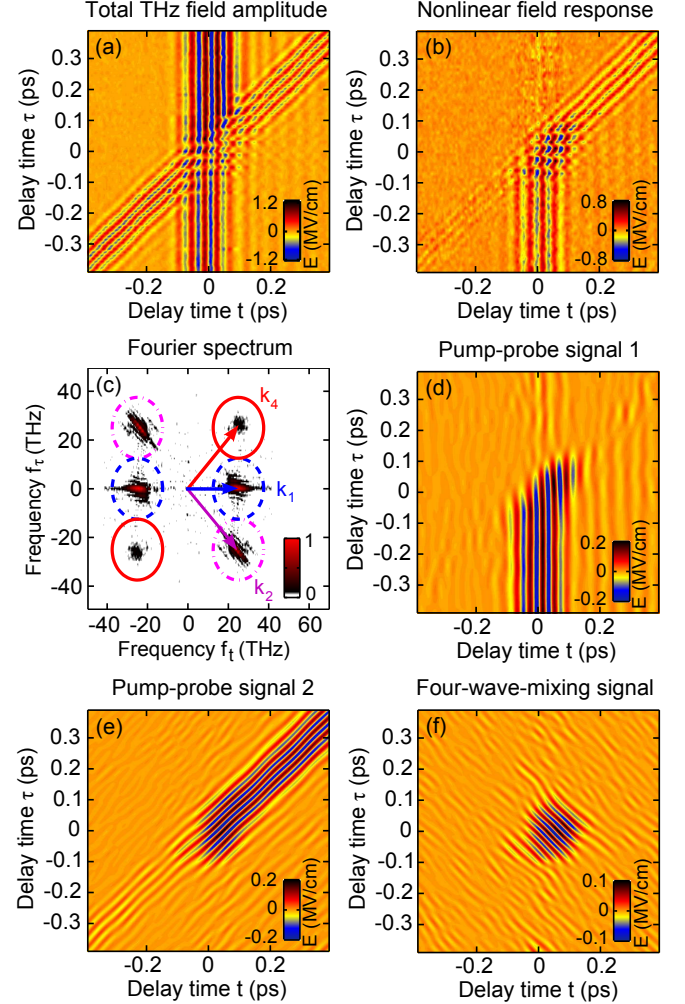}
\caption{\label{fig2} (a)~Electric field of two THz transients traversing InSb plotted as a function of delay time $\tau$ and sampling time $t$.
(b)~Nonlinear signal $E_{\text{NL}}$. (c)~2D FT of b) revealing pump-probe signals at wave vector positions $\bf{k_1}$ and $\bf{k_2}$ and the
FWM signature at $\bf{k_4}$. (d)~Selective inverse FT of the pump-probe signal emerging in the direction of $\bf{k_1}$. (e)~Inverse FT of the
pump-probe signature at $\bf{k_2}$ only. (f)~Inverse FT of the FWM signal in the direction of $\bf{k_4}$ only.}
\end{figure}

We use a two-dimensional scheme of nonlinear spectroscopy (Fig.~1(d)). In order to perform a THz multi-wave mixing experiment, a pair of
mutually synchronized pulses is obtained by splitting the THz beam after the GaSe emitter crystal. In this way nonlinear mixing effects of two
pulses within the emitter itself are precluded. A germanium (Ge) beam splitter set at Brewster's angle and coated with a 6-nm-thick gold layer
provides a splitting ratio of 1:1. The transmitted electric field in branch~2 ($E_2$) is delayed by a retroreflector stage. The transient in
branch~1 ($E_1$) propagates through a second Ge wafer to match the dispersion of $E_2$. The THz beams in both branches are individually chopped
with frequencies of 500~Hz and 250~Hz, respectively, and the chopper phases are locked to the 1~kHz pulse train. This configuration enables a
fast and efficient acquisition of the transmitted signals from only branch 1, only branch 2 or both branches at the same time ($E_{12}$). The
THz beams are tightly focused onto the same spot on the sample (FWHM: 85~$\mu$m). The emerging linear and nonlinear fields are collected with
large numerical aperture and directed onto a 140-$\mu$m-thick GaSe electro-optic sensor gated by near-infrared pulses with a duration of 8~fs.

Fig.~2(a) shows the total transmitted field $E_{12}$ as a function of the electro-optic sampling delay time $t$ and the relative temporal offset
$\tau$ between the THz pulses. The signal from branch~1 with a fixed temporal position appears as a set of vertical lines of constant phase
centered around $t = 0$~ps, whereas the diagonal lines correspond to the delayed signal from branch~2. The external peak fields are 2~MV/cm per
pulse. The field strengths within InSb are attenuated by a Fresnel factor $\tilde{t}=2/(n+1)=0.4$ defined by a refractive index of $n=4$ in the
frequency range of interest \cite{Adachi:87}. The nonlinear signal $E_{\text{NL}}$ shown in Fig.~2(b) is retrieved by subtracting the
contributions of individual transients $E_{1}$ and $E_{2}$ from the total response $E_{12}$: $E_{\text{NL}}=E_{12}-E_{1}-E_{2}$. The Fourier
transform of $E_{\text{NL}}$ (Fig.~2(c)) has a direct correspondence with the wave vector space and, thus, allows to disentangle different
contributions to the total nonlinear field response \cite{Kuehn:09}. The inverse Fourier transform of selected regions in frequency space
depicted in Fig.~2(c) provides the temporal fingerprints of multi-wave mixing signals of various orders. The pump-probe signals for each THz
pulse, located around wave vectors $\pm \bf{k_1}$ ($f_t=\pm f_0$, $f_{\tau}=0$) and $\pm \bf{k_2}$ ($f_t=\pm f_0$, $f_{\tau}=\mp f_0$)
(Fig.~2(c)), are shown in Figs.~2(d) and 2(e), respectively. These signals correspond to the transmission change of the sample excited by the
first and probed by the second THz transient. They do not depend on the relative phase.

Most remarkably, a FWM signal at the wave vector $\pm \bf{k_4}=\pm (2\bf{k_1}-\bf{k_2})$ ($f_t= \pm f_0$, $f_{\tau}= \pm f_0$) becomes clearly
discernible (Fig.~2(c)). This signal features lines of constant phase that point along a downward diagonal (Fig.~2(f)) indicating its dependence
on the respective phases of both THz transients. Thus, it bears information about coherence induced in the sample. Therefore, we will
concentrate in the following on the FWM signal which is free from influence of incoherent excitation processes dominating the pump-probe
response such as two-photon absorption \cite{Olszak:10} or impact ionization \cite{Hoffmann:09}.
\begin{figure}[b]
\includegraphics[width=8.6cm]{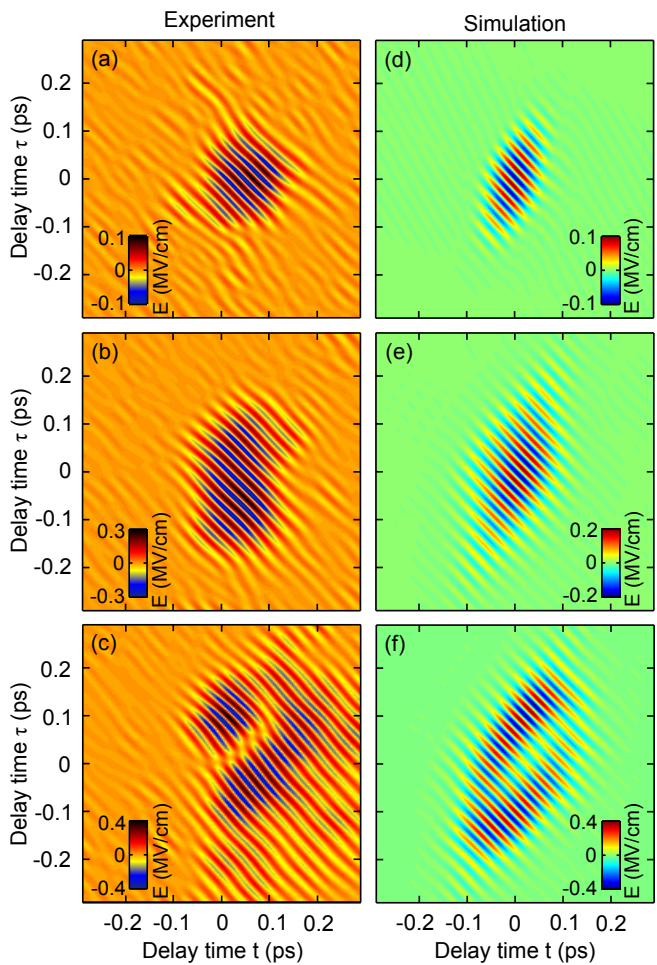}
\caption{\label{fig3} (a)~Oval FWM signal by applying an external field of 2~MV/cm per pulse. (b)~S-shaped FWM signature driven by an external
field of 3.5~MV/cm. (c)~Splitted FWM signal at an external exciting field strength of 5.3~MV/cm. (d-f)~Calculated FWM signatures reproducing the
main features of the corresponding measured signals on the left side of each image.}
\end{figure}

To study the field dependence of the FWM signal, external peak fields of 2~MV/cm, 3.5~MV/cm and 5.3~MV/cm per pulse are selected. The
corresponding pulse intensities are $I_0$, $3I_0$ and $7I_0$ where $I_0 = 10.6$~GW/cm$^2$. For the lowest peak electric field we observe an
oval-shaped envelope of the FWM signal, identical to the cross-correlation function of both pulses (Fig.~3(a)). This result is as expected in
the limit of perturbative nonlinear optics far from resonance. In contrast, increasing the field strength up to 3.5~MV/cm leads to a deviation
from the symmetric profile resulting in an S-shaped signal (Fig.~3(b)). Surprisingly, the maximum field of 5.3~MV/cm leads to a splitting of the
FWM signal (Fig.~3(c)). A minimum appears in the temporal region where the strongest total excitation field is present. This signature is an
unequivocal indication of an extremely nonlinear interaction.

Our experimental results may be understood by using a simplified model of a two-level system representing the interband resonance in InSb. The
simulation is based on the Maxwell-Bloch equations \cite{Ziolkowski:95} which are solved numerically without applying the
slowly-varying-envelope and rotating-wave approximations. The solution is obtained by an iterative predictor-corrector finite element method.
For the simulations we assume a dephasing time of $T_2=1$~ps \cite{Andrews:98} and a depopulation time of $T_1=10$~ps. Owing to the extremely
short THz transients the choice of the relaxation time barely affects the results of the simulations as long as $T_1$ and $T_2$ are longer than
the duration of the THz pulse. The transition dipole moment $\mu_{12} = 2.4$~e\AA\ and the density of the two-level systems $N=2.9 \times
10^{20}$~cm$^{-3}$ were adjusted in order to provide the best agreement with the shapes and intensities of the FWM signals measured
experimentally.
\begin{figure}[b]
\includegraphics[width=8.6cm]{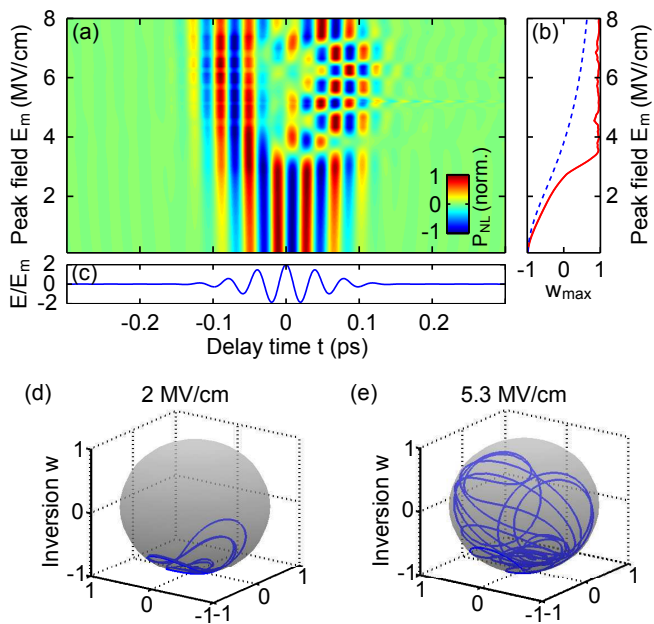}
\caption{\label{fig4} Simulated response of the two-level system for $\tau = 0$~ps: (a)~Normalized polarization of the two-level system at the
fundamental frequency of the driving field as a function of the delay time $t$ and the peak electric field $E_\text{m}$. (b)~Maximum inversion
$w_\text{max}$ as a function of $E_\text{m}$ (red solid line) and of the amplitude of a continuous wave excitation (blue dashed line)
(c)~Driving field of the THz transients. (d,e)~Corresponding pathways of the Bloch vector for a moderate ($E_\text{m}=2$~MV/cm) and a high
($E_\text{m}=5.3$~MV/cm) peak electric field.}
\end{figure}

Figs.~3(d)-(f) show the time domain FWM signals simulated for the same peak fields as those used for the experimental results depicted in the
panels on the left. For the simulation we assume Fourier-limited Gaussian THz transients with a FWHM duration set to the pulse width measured
experimentally. A perturbative response at moderate field strengths leads to an oval envelope of the FWM signal (Fig.~3(d)). At the intermediate
THz intensity this envelope evolves into an S-shape (Fig.~3(e)). Finally, external peak fields above 5~MV/cm induce a minimum in the center of
the signal, where the total field is strongest. This results in two side-lobes similar to those observed in the experiment (see Figs.~3(c) and
3(f)). As one can clearly see, the simulations with a single energy electronic resonance allow us to reproduce the essential features observed
in the experiment. This result is highly surprising since it is well known that interband excitations in bulk semiconductors like InSb include a
continuum of electronic resonances with a broad distribution of frequencies above the absorption edge. Nevertheless, simulations using a set of
two-level systems with a density distribution fitting the joint density of states in InSb essentially lead to the same results as those shown in
Figs.~3(c) and 3(f). The reason for that is the steep dependence of the non-perturbative response on the detuning frequency. This finding is
supported by the density of the two-level systems $N$ estimated from our simulations which constitutes only about 1\% of the total number of
available states in the conduction band of InSb. These are the states near the band edges which constitute the interband transitions with the
smallest detuning.

A qualitative physical understanding of the splitting observed in the FWM signal can be obtained by considering the polarization response of a
two-level system driven by two THz transients with zero delay time ($\tau=0$~ps). Fig.~4(a) depicts the simulated time-resolved polarization at
the frequency of the driving pulses for different peak electric fields $E_\text{m}$. In case of moderate fields the maximum population inversion
$w_\text{max}$ shown in Fig.~4(b) remains negative and the deflection of the Bloch vector from the ground state is minor as illustrated in
Fig.~4(d). Therefore, the response of the two-level system is perturbative and the shape of the polarization signal (Fig.~4(a)) follows the
profile of the driving fields shown in Fig.~4(c). The FWM signal, thus, can be described in terms of an instantaneous third-order nonlinearity.
However, this picture breaks down as soon as field amplitudes exceed 3~MV/cm and a clear splitting of the temporal signature of the FWM signal
starts to develop. In this case the maximum population inversion shown by the solid red line in Fig.~4(b) becomes positive ($w_\text{max} > 0$),
indicating the onset of a strongly non-perturbative regime. As illustrated in Fig.~4(e), a THz peak field of 5.3~MV/cm promotes the system
almost to the limit of complete population inversion. This surprising result radically differs from the case of an off-resonant excitation by a
continuous wave where the complete population inversion can be achieved only in the limit of infinitely high electric fields (see the dashed
blue line in Fig.~4(e)). This fact indicates a clear violation of the slowly-varying-envelope approximation for our conditions. Finally, in the
center of the FWM signal, where the driving electric field reaches its maximum, the polarization oscillates mainly at high harmonic frequencies
\cite{Tritschler:03a}. The response at the fundamental frequency becomes weak leading to the observed minimum (Fig.~4(a)). This regime of a
non-perturbative excitation sets in when the maximum Rabi frequency $\Omega_R/2\pi = 2\tilde{t}\mu_{12}E_m/h$ becomes comparable to the large
detuning of 18~THz at $E_m=3$~MV/cm.

In conclusion, we have studied the off-resonant FWM response of bulk InSb which provides a direct access to the coherent dynamics of the
interband polarization response at THz frequencies. The observed splitting of the FWM signals for electric fields above 5~MV/cm manifest the
onset of a non-perturbative response of Rabi flopping. This extremely nonlinear behavior underpins the high potential of the novel high-field
multi-THz technology for high harmonics generation \cite{Golde:08} and coherent control of quantum states in semiconductors
\cite{Leinss:08,Cole:01}. The high fields and excellent signal-to-noise performance of THz multi-wave mixing open a way for investigations of
the coherent response in a large variety of important resonances in complex systems such as intermolecular librations in hydrogen-bonded liquids
or energy gaps in superconducting condensates.

\end{document}